\documentclass[sigconf]{acmart}

\usepackage{booktabs}
\usepackage{algorithm}
\usepackage{algorithmic}
\usepackage{amsmath}
\usepackage{dblfloatfix}
\usepackage{balance}

\usepackage{dialogue}
\usepackage{tikz}
\usetikzlibrary{shapes.geometric, arrows}
\tikzstyle{item} = [rectangle, minimum width=3cm, minimum height=1cm,text centered, draw=black, fill=gray!50]
\tikzstyle{arrow} = [thick,->,>=stealth,shorten <= 1.5pt,shorten >= 1.5pt]

\renewcommand\footnotetextcopyrightpermission[1]{}

\setcopyright{none}

\acmYear{2020}

\acmConference{arXiv}{June 23}{2020}

\title{A Qualitative Evaluation of Language Models on Automatic Question-Answering for COVID-19}

\author{David Oniani}
\affiliation{%
  \institution{Mayo Clinic}
  \department{Kern Center for the Science of Health Care Delivery}
  \city{Rochester}
  \state{MN}
  \country{USA}}
\email{oniani.david@mayo.edu}

\author{Yanshan Wang}
\affiliation{%
  \institution{Mayo Clinic}
  \department{Division of Digital Health Sciences}
  \city{Rochester}
  \state{MN}
  \country{USA}}
\email{wang.yanshan@mayo.edu}

\begin{document}

%%%%%%%%%%%%%%%%%%%%%%%%%%%%%%%%%%%%%%%%%%%%%%%%%%%%%%%%%%%%%%%%%%%%%%%%%%%%%%%
% Abstract
%%%%%%%%%%%%%%%%%%%%%%%%%%%%%%%%%%%%%%%%%%%%%%%%%%%%%%%%%%%%%%%%%%%%%%%%%%%%%%%

\begin{abstract} \label{abstract}

COVID-19 (2019 Novel Coronavirus) has resulted in an ongoing pandemic and as of
12 June 2020, has caused more than 7.4 million cases and over 418,000 deaths.
The highly dynamic and rapidly evolving situation with COVID-19 has made it
difficult to access accurate, on-demand information regarding the disease.
Online communities, forums, and social media provide potential venues to search
for relevant questions and answers, or post questions and seek answers from
other members. However, due to the nature of such sites, there are always a
limited number of relevant questions and responses to search from, and posted
questions are rarely answered immediately. With the advancements in the field
of natural language processing, particularly in the domain of language models,
it has become possible to design chatbots that can automatically answer
consumer questions. However, such models are rarely applied and evaluated in
the healthcare domain, to meet the information needs with accurate and
up-to-date healthcare data. In this paper, we propose to apply a language model
for automatically answering questions related to COVID-19 and qualitatively
evaluate the generated responses. We utilized the GPT-2 language model and
applied transfer learning to retrain it on the COVID-19 Open Research Dataset
(CORD-19) corpus. In order to improve the quality of the generated responses,
we applied 4 different approaches, namely tf-idf (Term Frequency - Inverse
Document Frequency), Bidirectional Encoder Representations from Transformers
(BERT), Bidirectional Encoder Representations from Transformers for Biomedical
Text Mining (BioBERT), and Universal Sentence Encoder (USE) to filter and
retain relevant sentences in the responses. In the performance evaluation step,
we asked two medical experts to rate the responses. We found that BERT and
BioBERT, on average, outperform both tf-idf and USE in relevance-based sentence
filtering tasks. Additionally, based on the chatbot, we created a user-friendly
interactive web application to be hosted online and made its source code
available free of charge to anyone interested in running it locally, online, or
just for experimental purposes. Overall, our work has yielded significant
results in both designing a chatbot that produces high-quality responses to
COVID-19-related questions and comparing several embedding generation
techniques.

\end{abstract}


\maketitle

%%%%%%%%%%%%%%%%%%%%%%%%%%%%%%%%%%%%%%%%%%%%%%%%%%%%%%%%%%%%%%%%%%%%%%%%%%%%%%%
% Introduction
%%%%%%%%%%%%%%%%%%%%%%%%%%%%%%%%%%%%%%%%%%%%%%%%%%%%%%%%%%%%%%%%%%%%%%%%%%%%%%%

\section{Introduction} \label{introduction}

Coronavirus disease 2019 (COVID-19) is an infectious disease caused by severe
acute respiratory syndrome coronavirus 2 (SARS-CoV-2)~\cite{covid-19-cause}. As
of 12 June 2020, more than 7.4 million cases have been recorded, resulting in
over 418,000 deaths~\cite{covid-19-who-report}. The sudden global outbreak of
COVID-19 made millions of people quarantined, due to the social distancing
measures. Additionally, the COVID-19 pandemic caused a historic rise in mental
health problems, such as depression, post-traumatic stress disorder, and
suicide, due to the state-wise quarantine. People are isolated and stressed,
and may develop long-term psychological consequences, beyond the quarantine
period~\cite{covid-19-anxiety}~\cite{covid-19-mental-illness}~\cite{covid-19-mental-science}.
Therefore, most of the time, people rely on online and web-based resources for
getting news and updates concerning COVID-19. Given that currently many web
sources do not hold the accurate information about the pandemic and the
misinformation campaigns are running rampant~\cite{covid-19-misinformation}, it
is critically important that people and patients receive accurate, up-to-date,
and useful information regarding COVID-19. Online communities, forums, and
social media provide potential venues to search for relevant questions and
answers, or post questions and seek answers from other members. However, due to
the nature of such sites, there are always a limited number of relevant
questions and responses to search from, and posted questions are rarely
answered immediately. To address these issues, we propose to develop a chatbot
enhanced by neural language models that is able to automatically answer
questions related to COVID-19 through conversational interactions.

A conversational chatbot is a software which is able to conduct a conversation
via text and/or other means. There are different taxonomies for the type of
conversational chatbot. Based on how the natural language conversations are
generated, there are two main categories: script chatbot and intelligent
chatbot. The entire interaction in a script chatbot is based on a
pre-determined model that determines what the chatbot can and cannot do. The
``script'' is usually a decision tree that is manually crafted by domain
experts to determine which specific path to take given a response to one
question task. It is usually very labor-expensive and nongeneralizable to
develop conversation decision trees. The intelligent chatbot is built using
Artificial Intelligence (AI) and Natural Language Processing (NLP)
techniques that automatically generate
natural language on the back end. With the advancements in AI and NLP, the
functionality and the performance of modern chatbots have been dramatically
improved. However, these techniques are rarely applied and evaluated in the
healthcare domain to meet the information needs with accurate, up-to-date,
and interactive healthcare information.

The outbreak of COVID-19 has motivated us to develop a chatbot with advanced
NLP techniques and evaluate the approach in automatically answering questions
related to COVID-19. To the best of our knowledge, this is the first study of
such kind. Our contributions are:

\begin{itemize}
  \item We applied and compared the performance of four embedding generation
    approaches, namely tf-idf (Term Frequency - Inverse Document Frequency)~\cite{tf-idf},
    Bidirectional Encoder Representations from Transformers (BERT)~\cite{turc2019},
    BioBERT (Bidirectional Encoder Representations from Transformers for Biomedical
    Text Mining)~\cite{btz682}, and Universal Sentence Encoder (USE)~\cite{use}
    for refining the automatically generated answers.
  \item We proposed a qualitative evaluation guideline for automatic question-answering
    for COVID-19.
  \item We assessed the performance of the proposed ``hybrid'' approach for
    automatic question-answering for COVID-19.
  \item We built a web-based chatbot using the language models that facilitate
    question-answering for users.
\end{itemize}

This paper is organized as follows. We will proceed by discussing the related
work and the efforts in Section \ref{related}. Section \ref{materials} will be
dedicated to materials and Section \ref{methods} to the proposed approach. We
will report the chatbot evaluation strategy and the experimental results in
Sections \ref{discussion} and \ref{results}, respectively. Finally, we will
also discuss a web-based chatbot with the proposed model and future work in
Section \ref{discussion}, and conclude the work in Section \ref{conclusion}.

%%%%%%%%%%%%%%%%%%%%%%%%%%%%%%%%%%%%%%%%%%%%%%%%%%%%%%%%%%%%%%%%%%%%%%%%%%%%%%%
% Related Work
%%%%%%%%%%%%%%%%%%%%%%%%%%%%%%%%%%%%%%%%%%%%%%%%%%%%%%%%%%%%%%%%%%%%%%%%%%%%%%%

\section{Related Work} \label{related}

Recent neural language models of dialogue generation offer great promise for
generating responses for intelligent chatbots. The LSTM (Long Short-Term Memory)
sequence-to-sequence (seq2seq) model is one type of neural generation model that
maximizes the probability of generating a response given the previous dialogue
turn~\cite{long-short-term-memory}~\cite{lstm-question-answering}~\cite{seq2seq}.
XLNet uses a context of the word for predicting the next word where the context
word is constrained to two directions (backward or forward)~\cite{xlnet}. SAM
is a technique (Self-Attentive Associative Memory) where two memories are wired
into a single sequential model capable of both memorization and relational
reasoning~\cite{sam}.

In the GPT-2 domain, Lee and Hsiang~\cite{lee-hsiang} have fine-tuned GPT-2 for
generating patent claims. Klein and Nabi~\cite{klein-nabi} have applied GPT-2
in conjunction with BERT for automatic question generation purposes. 
Zhang, Sun, et al. developed a large and tunable neural conversational model
DialoGPT using GPT-2~\cite{zhang2019dialogpt}. Lee, Shu et al. developed RecipeGPT
for automatic generation of cooking recipes by fine-tuning GPT-2 on a large cooking
recipe dataset~\cite{recipegpt}. We are unaware of the work which applied GPT-2
model for transfer learning purposes on CORD-19.

In regard to the work related to comparing pretrained AI models, Jin et al.
made some efforts conducting probing experiments and comparing BERT, ELMo~\cite{elmo},
and BioBERT. Sharma and Daniel~\cite{bioflair} compared the performance of BERT
networks to that of FLAIR~\cite{akbik2018coling}.

In the general AI-based chatbot domain, Serbal et al.
~\cite{reinforce-chatbot} have applied deep reinforcement learning for building a
conversational AI chatbot. Adiwardana et al.~\cite{toward-human-chatbot} have
developed a multi-turn open-domain chatbot trained end-to-end on data mined social
media conversations. Yin et al.~\cite{therapy-chatbot} have developed a deep learning
based chatbot for psychological therapy purposes.

Semantic similarity of texts, on the other hand, has been studied for a long time
and recent breakthroughs allowed for development of new models such as BERT,
BioBERT, and Universal Sentence Encoder (USE). Today, one of the state-of-the art
conversational AI models is GPT-2. GPT-2 is a pretrained
model, so we have applied transfer learning utilizing CORD-19 for retraining
purposes. The resulted chatbot gave irregularly long responses that would not
be typical of a human. We have therefore decided to further filter the
responses via applying embedding generation algorithms and models such as
tf-idf, BERT, BioBERT, and USE and then using
semantic similarity approaches such as cosine similarity and inner product. In
other words, we first let a human ask a question and make GPT-2 come up with an
answer. We the further processed the response with additional filters and
ultimately, applied an embedding generation model for finding the sentences
that are most relevant to the question.

Cosine similarity is one of the most commonly used approaches in
calculating semantic similarity of texts. Therefore, it is naturally employed
in NLP tasks. Many NLP applications need to compute the semantic similarity
between two short texts. Its flexibility allows one to apply it under virtually
any settings, as long as documents can be represented as vectors. Besides,
finding cosine similarity is usually not a time-consuming task and can be done
really quickly. Therefore, it is also commonly used for benchmarking
purposes~\cite{cosine-benchmark}.

Our study has produced a chatbot that is both performant and extensible.
Additional layer of filters have shown success in classifying sentences.
The chatbot is also able to be retrained and readjusted to the new data, in
case there are new discoveries or scientific achievements related to COVID-19.
Furthermore, chatbot responses have been annotated by medical experts
and the results were consistent across the annotators.

%%%%%%%%%%%%%%%%%%%%%%%%%%%%%%%%%%%%%%%%%%%%%%%%%%%%%%%%%%%%%%%%%%%%%%%%%%%%%%%
% Materials
%%%%%%%%%%%%%%%%%%%%%%%%%%%%%%%%%%%%%%%%%%%%%%%%%%%%%%%%%%%%%%%%%%%%%%%%%%%%%%%

\section{Materials} \label{materials}

The White House Office of Science and Technology Policy alongside with the
coalition of leading research groups has released a COVID-19 machine readable
dataset - COVID-19 Open Research Dataset (CORD-19)~\cite{cord19}. It consisted
of over 128,000 scholarly articles regarding COVID-19, SARS-CoV-2, and related
coronaviruses, including over 59,000 with full text, and called researchers
globally to develop text and data mining tools for finding answers to the
questions within this content in support of the ongoing COVID-19 response
efforts worldwide~\cite{whitehousecovid2020}.

We used CORD-19 to train a language model that would automatically
answer questions related to COVID-19. The chatbot would not only help improve
information acquisition, but also serve as a knowledge base for COVID-19. We
harvested the data from the initial \textit{commercial use subset} of CORD-19,
containing 9000 scholarly articles in the form of JSON files. We extracted the
abstract and the main body of the article from every JSON file, combined them
together, and used as a corpus for retraining the language model.

%%%%%%%%%%%%%%%%%%%%%%%%%%%%%%%%%%%%%%%%%%%%%%%%%%%%%%%%%%%%%%%%%%%%%%%%%%%%%%%
% Methods
%%%%%%%%%%%%%%%%%%%%%%%%%%%%%%%%%%%%%%%%%%%%%%%%%%%%%%%%%%%%%%%%%%%%%%%%%%%%%%%

\section{Methods} \label{methods}

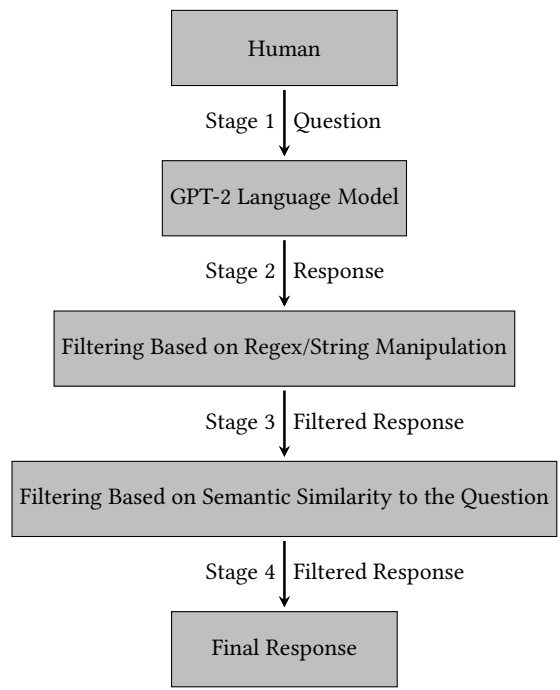
\begin{figure}%[H]
  \centering
  \begin{tikzpicture}[node distance=2cm]
    \node (human)        [item]                        {Human};
    \node (chatbot)      [item, below of=human]        {GPT-2 Language Model};
    \node (filtering)    [item, below of=chatbot]      {Filtering Based on Regex/String
      Manipulation};
    \node (filtering-ss) [item, below of=filtering]    {Filtering Based on Semantic
      Similarity to the Question};
    \node (final)        [item, below of=filtering-ss] {Final Response};
    
    \draw [arrow] (human)        -- node[anchor=east] {Stage 1} (chatbot);
    \draw [arrow] (chatbot)      -- node[anchor=east] {Stage 2} (filtering);
    \draw [arrow] (filtering)    -- node[anchor=east] {Stage 3} (filtering-ss);
    \draw [arrow] (filtering-ss) -- node[anchor=east] {Stage 4} (final);
    
    \draw [arrow] (human)        -- node[anchor=west] {Question}          (chatbot);
    \draw [arrow] (chatbot)      -- node[anchor=west] {Response}          (filtering);
    \draw [arrow] (filtering)    -- node[anchor=west] {Filtered Response} (filtering-ss);
    \draw [arrow] (filtering-ss) -- node[anchor=west] {Filtered Response} (final);
  \end{tikzpicture}
  \caption{Workflow for Response Generation.}
  \label{fig.workflow}
\end{figure}

We applied a hybrid approach for generating responses: GPT-2 was used to
generate the answer to the question, then an additional filtering step
was applied for pruning the irrelevant sentences from the answer, and
subsequently, semantic similarity methods were employed to retain the sentences
that are most semantically similar to the question. Such hybrid approach to the
response generation produced high quality answers to COVID-19-related
questions. Figure \ref{fig.workflow} illustrates the pipeline of the proposed
approach.

\subsection{GPT-2 Language Model}

GPT-2 has a Transformer-based~\cite{vaswani} architecture which, in
many ways, is similar to Open AI GPT model\cite{radford2019language}\cite{gpt}.

There are a total of 4 different GPT-2 models that were released by OpenAI:
124 million (124M), 355 million (355M), 774 million (774M), and 1.5 billion (1.5B)
parameters~\cite{github-openai} models. While the model with 1.5 billion parameters
showed the best results in the original paper~\cite{radford2019language}, in our
experiments, we found that it was difficult to fine-tune and use for the transfer
learning purposes. Besides, the training was unbearably slow, even if run on TPUs
(Tensor Processing Unit) provided by Google Colaboratory~\cite{google-colab} which
we used as our training ground.

We therefore utilized 774M model and ran transfer learning for 2500 iterations with the
batch size of 8. After 2000 iterations, the loss was not decreasing so we let
the language model train for the additional 500 iterations and stopped the training.
The batch size of 8 was chosen due to the memory limitations of Google Colaboratory.
As for the optimizer, we used Adam~\cite{adam} and set the learning rate of 0.0001 (\texttt{1e-4}).

Adam is an algorithm for first-order gradient-based optimization of
stochastic objective functions, based on adaptive estimates of lower-order
moments~\cite{adam}. It is highly memory-efficient and has shown good results
in retraining our chatbot. We have also tried SGD~\cite{SGD}, yet Adam has
shown the better performance and hence, we have released the Adam-based
retrained model.

The original GPT-2 was written in tensorflow~\cite{tensorflow2015-whitepaper}
and this is the version we used. That said, for retraining purposes, we applied
the TPU-trainable version of the GPT-2~\cite{gpt-2-tpu}.

As for the hardware, Google Colaboratory provided us with cloud TPUs and
training capabilities. It came 25 GB RAM and since we connected the Colab
to Google Drive~\cite{google-colab}, we had enough storage to do transfer
learning.

The link for downloading the model is available on our GitHub page~\cite{github-covid-19-chatbot}.

%%%%%%%%%%%%%%%%%%%%%%%%%%%%%%%%%%%%%%%%%%%%%%%%%%%%%%%%%%%%%%%%%%%%%%%%%%%%%%%
% Regex Filtering
%%%%%%%%%%%%%%%%%%%%%%%%%%%%%%%%%%%%%%%%%%%%%%%%%%%%%%%%%%%%%%%%%%%%%%%%%%%%%%%

\subsection{Filtering Based on Regex/String Manipulation}

The GPT-2 responses are usually very lengthy and for the most part,
the answer is not relevant to the question. To prune the responses generated
from GPT-2, we first chunked the answer into the list of sentences using
Python's built-in module for dealing with regular expressions (\texttt{re}~\cite{python-re})
and then for each answer in the list of answers, performed the following
regex/string operations:

\begin{enumerate}
  \item Eliminated redundant spaces
  \item Eliminated extra punctuation marks (specifically, ``.'', ``!'', and ``?'')
  \item Removed redundant parentheses and square brackets
  \item Further split the sentence into separate sentences if it contained a period (``.'')
\end{enumerate}

Steps 2 and 4, once again, employed \texttt{re} module while for steps 1 and 4,
just the built-in string operations were sufficient (hence, no built-in or external
module was used).

These operations have significantly improved the quality of the answer and allowed
us directly passing them to the pretrained models for generating embeddings.

%%%%%%%%%%%%%%%%%%%%%%%%%%%%%%%%%%%%%%%%%%%%%%%%%%%%%%%%%%%%%%%%%%%%%%%%%%%%%%%
% Semantic Similarity
%%%%%%%%%%%%%%%%%%%%%%%%%%%%%%%%%%%%%%%%%%%%%%%%%%%%%%%%%%%%%%%%%%%%%%%%%%%%%%%

\subsection{Filtering Based on Semantic Similarity to the Question}

Semantic similarity is a metric that quantifies the degree to which two texts
or text documents are similar to each other. The two approaches we have used
include cosine similarity and inner product. The difference between the two
is that cosine similarity pays attention to only the angle between the vectors,
while the inner product cares about both the angle and the magnitude. That said,
if one has the normalized data, both approaches are nearly equivalent.

To put each sentence in a vector representation, we tested and applied
four different approaches for generating embeddings:

\begin{itemize}
  \item tf-idf~\cite{tf-idf-scikit}: a simple, tf-idf based embedding-generation method.
  \item BERT: Large model, uncased (24 layers and 340M parameters)~\cite{github-bert}.
  \item BioBERT: Large v1.1 (+PubMed 1M) model based on BERT-large Cased (custom 30k
    vocabulary)~\cite{github-biobert}.
  \item USE: Version 3, large~\cite{tfhub-use}.
\end{itemize}

In all cases, the similar strategy was applied for filtering sentences. The
following equation defines the embedding generation process:

\begin{equation}
  E = emb(S)
\end{equation}

where $S = (s_1, s_2, \dots s_n, q)$ and denotes the list of sentences obtained
by performing the split based on regex/string manipulation plus the question,
$emb$ denotes one of the embedding generation approaches (i.e., tf-idf, BERT,
BioBERT, or USE), and $q$ denotes the question being asked. $E$ denotes the
embeddings generated for the list of sentences. In other words, the list of
sentences is passed to the embedding generation function and the feature matrix
is returned.

Once the embeddings were generated, we have applied cosine similarity and inner
product for ranking the sentences. Inner product was used only with USE since
its embeddings were approximately normalized. The following equation shows the
ranking process:

\begin{equation}
  M = (sim(e_1, e_q), sim(e_2, e_q) \dots sim(e_n, e_q))
\end{equation}

where $e_1, e_2, \dots e_n, e_q \in E$ and $sim$ is a function that finds the
cosine similarity or the inner product between two embeddings. For every
sentence embedding $e_1, e_2, \dots e_n$, the approach calculates the cosine
similarity to the question embedding $e_q$. $M$ is a similarity matrix where
every column denotes the similarity vector for a particular sentence. The
similarity vector contains the similarity values of the particular sentence
with respect to all the sentences (including itself). The last
column/similarity vector in the matrix has the similarity measures with respect
to the question embedding.

Finally, we ranked the sentences based on their similarity scores, selected top
five most similar sentences using Python's numpy~\cite{numpy} package, and
concatenated them into a single, unified answer.

%%%%%%%%%%%%%%%%%%%%%%%%%%%%%%%%%%%%%%%%%%%%%%%%%%%%%%%%%%%%%%%%%%%%%%%%%%%%%%%
% Example Responses
%%%%%%%%%%%%%%%%%%%%%%%%%%%%%%%%%%%%%%%%%%%%%%%%%%%%%%%%%%%%%%%%%%%%%%%%%%%%%%%

\subsection{Example Responses}

We show an example responses in the form of a Human-Bot
conversation generated by the GPT-2 language model, as follows. 

\begin{dialogue}
  \speak{Human} What do we know about vaccines and therapeutics of COVID-19?
  \speak{Bot} Because of their critical role in viral pathogenesis and host
      defense against infections, several drugs have been developed against COVID-19 With this in mind, we present evidence that drugs that inhibit pro-inflammatory responses (e.g., IL-6) have the potential to inhibit COVID-19 infection and viral replication in nonhuman primates after administration up to 12 days following infection Because of their critical role in viral pathogenesis and host defense against infections, several drugs have been developed against COVID-19
  \speak{Human} What do we know about COVID-19 risk factors?
  \speak{Bot} However, COVID-19 is now a threat to global health security
      According to the latest World Health Organization report, a large number of travel is seen as one of the major causes of COVID-19 spread However, COVID-19 is now a threat to global health security
  \speak{Human} What is known about transmission, incubation, and environmental stability of COVID-19?
  \speak{Bot} For example, a number of reports have suggested that animal
       and human hosts may have provided the necessary environment for the transmission of COVID-19 in people and poultry In fact, all known animal hosts that have been associated with animal-to-human transmission of COVID-19 are wild birds, including chickens, geese, mule deer, herons, and turkey In fact, all known animal hosts that have been associated with animal-to-human transmission of COVID-19 are wild birds, including chickens, geese, mule deer, herons, and turkey
\end{dialogue}

The dialogue shows a consistent pattern in generating responses.
Response to the first question regarding the vaccines and therapeutics states
that ``several drugs have been developed against COVID-19.'' Furthermore, it
says that ``drugs that inhibit pro-inflammatory responses (e.g., IL-6) have the
potential to inhibit COVID-19 infection and viral replication in nonhuman
primates.'' Regarding the response to the second question, concerning the risk
factors, it addressed the question directly by stating that ``a large number of
travel is seen as one of the major causes of COVID-19 spread.'' As for the
third question, about transmission, incubation, and environmental stability of
COVID-19, it has mentioned that ``large number of travel is seen as one of the
major causes of COVID-19 spread'' and additionally, talks about ``animal-to-human
transmission.'' In all cases, sentences were highly readable and understandable.
That said, in some cases, the same sentences were repeated due to how the hybrid
approach was implemented. This can be avoided, which we discuss in the section
\ref{discussion}.

%%%%%%%%%%%%%%%%%%%%%%%%%%%%%%%%%%%%%%%%%%%%%%%%%%%%%%%%%%%%%%%%%%%%%%%%%%%%%%%
% Questions and Evaluation
%%%%%%%%%%%%%%%%%%%%%%%%%%%%%%%%%%%%%%%%%%%%%%%%%%%%%%%%%%%%%%%%%%%%%%%%%%%%%%%

\section{Questions and Evaluation} \label{evaluation}

In order to evaluate the performance of the proposed approaches as well as the
overall performance of the chatbot, it is crucial to have a question dataset
that both are frequently asked and related to COVID-19. For this purpose, we
decided to use 12 questions from the Kaggle's COVID-19 Open Research Dataset
Challenge (CORD-19)~\cite{kaggle}. Most of the questions included the term
``COVID-19'' but others did not, in which case we appended the term to the
end of the question. Table \ref{tab.question} presents all 12 questions.

\begin{table}
\caption{Testing questions from CORD-19.}
\label{tab.question}
  \begin{tabular}{p{0.1\linewidth}p{0.7\linewidth}}
    \toprule
    Number & Question\\
    \midrule
    \#1 & Are there geographic variations in the mortality rate of COVID-19?\\
    \midrule
    \#2 & What is known about transmission, incubation, and environmental
      stability of COVID-19?\\
    \midrule
    \#3 & Is there any evidence to suggest geographic based virus mutations of
      COVID-19?\\
    \midrule
    \#4 & Are there geographic variations in the rate of COVID-19 spread?\\
    \midrule
    \#5 & What do we know about virus genetics, origin, and evolution of
      COVID-19?\\
    \midrule
    \#6 & What has been published about ethical and social science considerations
      of COVID-19?\\
    \midrule
    \#7 & What has been published about medical care of COVID-19?\\
    \midrule
    \#8 & What do we know about diagnostics and surveillance of COVID-19?\\
    \midrule
    \#9 & What do we know about COVID-19 risk factors?\\
    \midrule
    \#10 & What has been published about information sharing and inter-sectoral
      collaboration of COVID-19?\\
    \midrule
    \#11 & What do we know about vaccines and therapeutics of COVID-19?\\
    \midrule
    \#12 & What do we know about non-pharmaceutical interventions of COVID-19?\\
    \bottomrule
  \end{tabular}
\end{table}

For every one of the 12 questions, we generated five different answers by
applying four different embedding generation techniques,
resulting in a total of 240 answers. Therefore, the response for every question
was generated exactly 5 times using the same technique. This ensured a fair and
consistent distribution of both the questions and the approaches across the
dataset. We made all of the answers publicly available on
GitHub~\cite{github-results}. We then asked two experienced medical experts to
evaluate the quality of these responses by assigning different relevance scores
according to the categories in Table \ref{tb.score.annotation.categories}.
Having 5 categories allowed for a flexibility and diversity of
opinions/judgements as well as a broad range of scores that ultimately gave us
a better way to evaluate our approaches. The evaluation was done primarily by
averaging the scores for a particular approach.

\begin{table}
\caption{5 Rating Categories.}
  \label{tb.score.annotation.categories}
  \begin{tabular}{p{0.2\linewidth}p{0.5\linewidth}p{0.1\linewidth}}
    \toprule
    Category & Description & Point(s)\\
    \midrule
    Relevant & The answer partially or fully answers the question and/or
      makes clear attempts to do so and is related to the question & 5\\
    \midrule
    Well-formed & the answer makes a logical sense and is somewhat related
      to both the question and COVID-19, yet it does not (partially or 
      fully) answer the question & 4\\
    \midrule
    Informative & The answer is not related to the question, but provides
      some information about COVID-19 and makes a logical sense & 3\\
    \midrule
    Acceptable & The answer makes some logical sense and is weakly related
      to the question or COVID-19, but is mostly difficult to understand &
      2\\
    \midrule
    Poor & the answer is totally unrelated to the question or COVID-19
      and/or does not make a logical sense & 2\\
    \bottomrule
  \end{tabular}
\end{table}

Our annotation process had two phases. In the first phase, we let the
annotators evaluate the test subset of the responses generated by the language
model. The test subset was comprised of 20 questions. We then computed the
IAA (Inner Annotator Agreement) which was approximately equal to 0.389. Due to
having 5 categories, we used Pearson correlation coefficient for computing the
IAA (as opposed to Cohen's Kappa, etc). Low correlation value led us to having
a meeting with both annotators where we discussed why the they had different
scores on the particular responses to questions. Finally, both annotators reached
the agreement and gave same scores for every question in the test subset of 20.
Once the agreement was reached, we then let the annotators evaluate the
remaining 220 questions. Note that we evaluated our model based on the 240
responses and included initial subset, where both annotators agreed on the
judgement. This was done for the sake of fairness and consistency.

%%%%%%%%%%%%%%%%%%%%%%%%%%%%%%%%%%%%%%%%%%%%%%%%%%%%%%%%%%%%%%%%%%%%%%%%%%%%%%%
% Results
%%%%%%%%%%%%%%%%%%%%%%%%%%%%%%%%%%%%%%%%%%%%%%%%%%%%%%%%%%%%%%%%%%%%%%%%%%%%%%%

\section{Empirical Results} \label{results}

\subsection{Performance by Approach.}

Table \ref{tb.score.results} lists the evaluation results of different
approaches. It shows the approach, the average scores based on the approach for
each annotator, and the overall average across the annotators. The first
annotator rated BERT as the best approach with the average score of 4.167.
BioBERT shows slightly worse performance with a score of 4.133 than BERT. The
tf-idf approach performs well with a score of 3.967, yet it could not
outperform either BERT or BioBERT. USE has the worst performance out of all
embedding generation techniques with the score of 3.683 out of 5. The second
annotator, similarly, gave the highest average score to BERT (4.283). USE was
the second best with the score of 4.083 followed by BioBERT with approximately
the same score of 4.067. The tf-idf approach has yielded the worst results,
rated 3.8.

In general, the results are consistent between two annotators with an inner
annotator agreement score of 0.521, which was calculated using the Pearson
correlation. Models from the BERT family showed the best
performance in automatically answering  COVID-19 questions, with BERT slightly
outperforming BioBERT (4.225 vs. 4.100 - average scores) being the best. The
tf-idf approach and USE show roughly similar performance (3.884 vs. 3883)., yet
inferior to BERT and BioBERT. All four approaches, on average, can be
considered to be in the ``well-formed'' category with BERT and BioBERT being
close to the ``Relevant'' category. The overall average was 4.023
(Well-formed).

\begin{table}
  \caption{Average Scores of Embedding Generation Approaches Across the Annotators.}
  \label{tb.score.results}
  \begin{tabular}{p{0.2\linewidth}p{0.2\linewidth}p{0.2\linewidth}p{0.2\linewidth}}
    \toprule
    Approach & A1 & A2 & Overall\\
    \midrule
    tf-idf & 3.967 & 3.8 & 3.884\\
    \midrule
    BERT & 4.167 & 4.283 & \textbf{4.225}\\
    \midrule
    BioBERT & 4.133 & 4.067 & 4.100\\
    \midrule
    USE & 3.683 & 4.083 & 3.883\\
    \bottomrule
  \end{tabular}
\end{table}

\subsection{Performance by Question}

Table \ref{tb.score.question} shows the average scores for annotators A1 and
A2, the overall average, and the difference based on the question asked to the
language model. From the table, it is clear that the proposed approach had the
best results for responses to questions \# 11, \# 3, and \# 9 with average
relevance scores of 4.700, 4.700, and 4.550, respectively After conducting the
one-tailed one sample t-test, taking the differences as the values, we conclude
that the differences for the scores (0.200, 0.300, and 0.200 respectively) are
not significant with p-values being approximately 0.682, 0.221, and 0.682
respectively (all being above our threshold value of 0.05). This means that
both annotators made consistent judgement.

It should be noted that all 3 of these questions seem to be rather short in
length. The responses to question \# 7, on the other hand, has the worst
average score and interestingly, both annotators have given the same score of
2.667. That said, the question is also one of the shortest questions in length.
Therefore, the length does not seem to always have a correlation with the score
(hence, the performance).

To further analyze why responses to the question \# 7 had the lowest average
score, we determined whether the terms of the question are present in the
dataset. Terms ``inter-sectoral collaboration'' and ``information sharing''
were both present in CORD-19. Therefore, it is likely that the issue stems from
the model itself and not the dataset.

According to the scores, we also find that the terms in the question has some
correlation with the score. For example, the questions that featured words
strongly linked to COVID-19, such as virus in question \# 3, vaccine in
question \# 11, and risk in question \# 9, had higher average response scores
than those that did not (e.g., question \# 10).

\begin{table}
  \caption{Question-Based Average Scores of Embedding Generation Approaches Across the
    Annotators.}
  \label{tb.score.question}
  \begin{tabular}{p{0.16\linewidth}p{0.16\linewidth}p{0.16\linewidth}p{0.16\linewidth}p{0.16\linewidth}}
    \toprule
    Question & A1 & A2 & Average & Difference\\
    \midrule
    \# 1 & 4.200 & 3.450 & 3.825 & 0.750 (A1)\\
    \midrule
    \# 2 & 4.350 & 4.100 & 4.225 & 0.250 (A1)\\
    \midrule
    \# 3 & 4.550 & 4.850 & \textbf{4.700} & 0.300 (A2)\\
    \midrule
    \# 4 & 4.100 & 4.150 & 4.125 & 0.050 (A2)\\
    \midrule
    \# 5 & 3.600 & 3.950 & 3.775 & 0.350 (A2)\\
    \midrule
    \# 6 & 4.100 & 4.250 & 4.175 & 0.150 (A2)\\
    \midrule
    \# 7 & 2.650 & 2.650 & 2.650 & 0.000 (NA)\\
    \midrule
    \# 8 & 3.850 & 4.450 & 4.150 & 0.000 (A2)\\
    \midrule
    \# 9 & 4.650 & 4.450 & 4.550 & 0.200 (A1)\\
    \midrule
    \# 10 & 3.412 & 3.706 & 3.559 & 0.294 (A2)\\
    \midrule
    \# 11 & 4.600 & 4.800 & \textbf{4.700} & 0.200 (A2)\\
    \midrule
    \# 12 & 3.700 & 3.850 & 3.775 & 0.150 (A2)\\
    \bottomrule
  \end{tabular}
\end{table}

%%%%%%%%%%%%%%%%%%%%%%%%%%%%%%%%%%%%%%%%%%%%%%%%%%%%%%%%%%%%%%%%%%%%%%%%%%%%%%%
% Discussion
%%%%%%%%%%%%%%%%%%%%%%%%%%%%%%%%%%%%%%%%%%%%%%%%%%%%%%%%%%%%%%%%%%%%%%%%%%%%%%%

\section{Discussion} \label{discussion}

%%%%%%%%%%%%%%%%%%%%%%%%%%%%%%%%%%%%%%%%%%%%%%%%%%%%%%%%%%%%%%%%%%%%%%%%%%%%%%%
% Limitations
%%%%%%%%%%%%%%%%%%%%%%%%%%%%%%%%%%%%%%%%%%%%%%%%%%%%%%%%%%%%%%%%%%%%%%%%%%%%%%%

The project had several limitations. First, due to hardware constraints and the
difficulty of fine-tuining, we have not used the larger 1.5B GPT-2 model that
could potentially yield better results in generating responses. Second, the
question pool was also limited and comprised of 12 questions. Additionally, we
have tried only 4 specific embedding generation approaches, which might not be
a fair representation of all such techniques in the domains of AI and NLP.

%%%%%%%%%%%%%%%%%%%%%%%%%%%%%%%%%%%%%%%%%%%%%%%%%%%%%%%%%%%%%%%%%%%%%%%%%%%%%%%
% Web Application
%%%%%%%%%%%%%%%%%%%%%%%%%%%%%%%%%%%%%%%%%%%%%%%%%%%%%%%%%%%%%%%%%%%%%%%%%%%%%%%

In order to make the language model more accessible to the general audience for
automating the response generation, we built a web-based chatbot using the
trained GPT-2 with options of tf-idf, BERT, BioBERT, and USE approaches. Please
find the released code on our GitHub:
\href{https://github.com/oniani/covid-19-chatbot/tree/master/web-app}. The
application is powered by Python's Flask~\cite{flask} package and gives a
simple and user-friendly interface for the interactive communication with the
chatbot. Please note that the health information generated by the chatbot is
for general research purposes only. It is not a diagnostic tool, nor is it a
substitute for medical advice or treatment for specific conditions.

%%%%%%%%%%%%%%%%%%%%%%%%%%%%%%%%%%%%%%%%%%%%%%%%%%%%%%%%%%%%%%%%%%%%%%%%%%%%%%%
% Future Work
%%%%%%%%%%%%%%%%%%%%%%%%%%%%%%%%%%%%%%%%%%%%%%%%%%%%%%%%%%%%%%%%%%%%%%%%%%%%%%%

Although our work has demonstrated the feasiblity of using language models for
automatically answering COVID-19 questions, much can be done in the further
research. At first, we would like to explore why certain questions had the
higher scores than others. Secondly, other approaches for generating embeddings
for sentences, such as BioWordVec~\cite{biowordvec}, could potentially improve
the performance of the chatbot and can be another avenue for exploration. From
the dialogue presented in Section \ref{methods}, it is clear that GPT-2 could
generate duplicate sentences that may be irrelevant to the question. In that
case, the same sentence might be repeated in the final answer. One could
incorporate an additional simple step of eliminating duplicate sentences which
could potentially improve the quality of the answers. Adding an additional,
third layer of filtering can also be tested and checked whether it improves the
quality of the responses. Additionally, the GPT-2 model can always be further
retrained on a new corpus, which could potentially improve the result. The 1.5B
GPT-2 model could also be applied for retraining purposes. Finally, given that
a larger GPT language model (GPT-3) was recently released
~\cite{brown2020language}, we believe that it is feasible for the chatbot with
the capable hardware to explore the realm of possibilities with this model
which could also evolve into an interesting future work.

%%%%%%%%%%%%%%%%%%%%%%%%%%%%%%%%%%%%%%%%%%%%%%%%%%%%%%%%%%%%%%%%%%%%%%%%%%%%%%%
% Conclusion
%%%%%%%%%%%%%%%%%%%%%%%%%%%%%%%%%%%%%%%%%%%%%%%%%%%%%%%%%%%%%%%%%%%%%%%%%%%%%%%

\section{Conclusion} \label{conclusion}

In this paper, we applied the GPT-2 language model to automatically answer
questions related to COVID-19, and quantitatively evaluate the proposed
approach. To refine the responses generated by GPT-2, we compared four
different embedding generation techniques, namely tf-idf, BERT, BioBERT, and
USE. We utilized the collected corpus from the CORD-19 task to pretrain the
GPT-2 model, and evaluated the automatically generated answers on twelve
questions from the CORD-19. The results were evaluated by two medical experts.
In general, the results are consistent between two annotators. The empirical
results show that  BERT achieved the best performance in automatically
answering  COVID-19 questions. We also built a web-based chatbot using the
trained GPT-2 model and opensoured the code.

\begin{acks}
This work was supported by NIH grant R01LM11934, the Mayo Clinic Center for
Health Equity and Community Engagement Research Award, and the Mayo Clinic
Office of Patient Education. The funders had no role in the design of the
study, or collection, analysis, and interpretation of data and in preparation
of the manuscript. The views presented in this report are not necessarily
representative of the funder's views and belong solely to the authors.
\end{acks}

%%%%%%%%%%%%%%%%%%%%%%%%%%%%%%%%%%%%%%%%%%%%%%%%%%%%%%%%%%%%%%%%%%%%%%%%%%%%%%%
% References
%%%%%%%%%%%%%%%%%%%%%%%%%%%%%%%%%%%%%%%%%%%%%%%%%%%%%%%%%%%%%%%%%%%%%%%%%%%%%%%

\bibliographystyle{ACM-Reference-Format}
\balance
\bibliography{paper}

\end{document}